\documentclass[preprint,12pt]{elsarticle}

\usepackage{amssymb,amsmath}
\usepackage{hyperref}

\begin{document}

\begin{frontmatter}

\title{Generalization properties of restricted Boltzmann machine for short-range order}

\author{M. A. Timirgazin}
\author{A. K. Arzhnikov}

\affiliation{organization={Physical-Technical Institute, UdmFRC UB RAS},
            addressline={T. Baramzinoy st., 34}, 
            city={Izhevsk},
            postcode={426067}, 
            state={Udmurt Republic},
            country={Russia}}

\begin{abstract}
The restricted Boltzmann machine (RBM) is used to investigate short-range order in binary alloys. The network is trained on the data collected by Monte Carlo simulations for a simple Ising-like binary alloy model and used to calculate the Warren--Cowley short-range order parameter and other thermodynamic properties. We demonstrate that RBM not only reproduces the order parameters for the alloy concentration at which it was trained, but can also predict them for any other concentrations.
\end{abstract}




\end{frontmatter}


\section{\label{sec:intro}Introduction}

The use of machine learning (ML) methods has proved its efficiency in various technical devices. Neural networks have made a breakthrough in speech and image recognition technologies, the implementation of control of complex technical devices and manufacturing processes. In recent years, ML has grown in popularity as a powerful scientific tool. Neural networks are used for identification and classification of phases in classical and quantum systems~\cite{Wang2016,Carrasquilla2017,Hu2017,Shiina2020,Wetzel2017,Chng2017,Westerhout2020},
accelerating Monte Carlo~\cite{Liu2017,Huang2017,Shen2018} and molecular dynamics simulations~\cite{Noe2020,Smith2017}, and many other fields (see the reviews~\cite{Carleo2019,Mehta2019}).

Among the various architectures of neural networks, of particular interest is the restricted Boltzmann machine (RBM), a simple generative energy-based model~\cite{Ackley1985}. The main feature of RBM is the hidden layer which in physical language can be interpreted as an introduced auxiliary field that allows you to decouple complex interaction between visible variables (one can draw an analogy with the Hubbard--Stratonovich transformation~\cite{Stratonovich1957,Rrapaj2020}). Based on principles of statistical mechanics RBM turns out to be able to capture the physics underlying classical and quantum many-body problems~\cite{Torlai2016,Nomura2017,Rrapaj2020,Vieijra2020}. It's important that RBM can not only classify and process large data, but can be used even for reconstructing the ground state wave function~\cite{Carleo2017}. While RBM usually is treated as a component of deep belief networks~\cite{Hinton2006}, it is found that at least for Ising model the shallow RBM gives just as good results as deep models being much more simple in training~\cite{Morningstar2017}.

While neural networks prove their ability to effectively calculate physical properties being trained on the preliminary prepared data set, their generalization properties, i.~e. ability to adapt properly to new data and make correct predictions, are still under the question. We train RBM on a binary system with the short-range order and demonstrate how the network can not only reproduce order parameters for the concentration at which it was trained but predict order parameters for any other concentration.

\begin{figure}[h]
\includegraphics[width=0.8\textwidth]{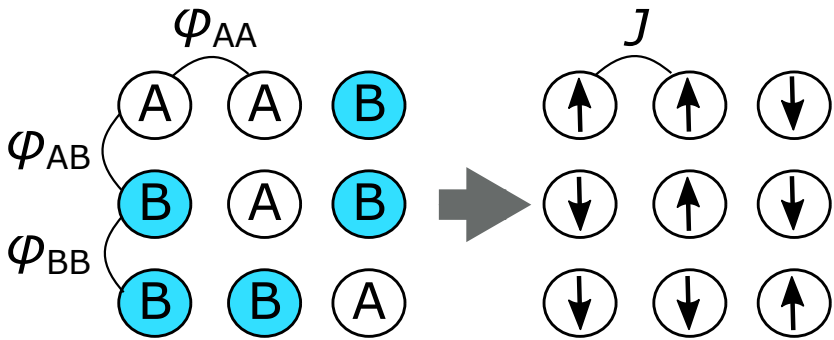}
\caption{\label{fig:binary}Transformation of binary (A and B) alloy problem to Ising spin model on square lattice.}
\end{figure}

\section{\label{sec:model}Model}
\subsection{\label{sub:SRO}Short-range order}
Short-range order (SRO) refers to the regular and predictable arrangement on a local length scale.
A simple system which manifests SRO is the binary alloy with type-$A$ and type-$B$ atoms (Fig.~\ref{fig:binary})~\cite{Ziman1979}. The crystal energy of the binary alloy can be represented as a sum of pair potentials:
\begin{equation}
E=N_{AA} \phi_{AA}+N_{BB} \phi_{BB}+N_{AB} \phi_{AB}, \label{eq:ham0}
\end{equation}
where $N_{\alpha\beta}$ is the number of atom pairs $\alpha$-$\beta$, $\phi_{\alpha\beta}$ is their energy.  Let the variable $S_i$ take the value $+1$ or $-1$ if the site $i$ occupied by an atom $A$ or $B$, respectively. Then the energy of the system (\ref{eq:ham0}) can be written in the Ising-like form~\cite{Ziman1979}:
\begin{equation}
\mathcal{H}=-J\sum_{\langle i,j\rangle}S_i S_j - h\sum_i S_i+C_0, \label{eq:ham1}
\end{equation}
where $J=\frac{1}{2}[\phi_{AB}-\frac{1}{2}(\phi_{AA}+\phi_{BB})]$, $h=\frac{1}{4}z(\phi_{BB}-\phi_{AA})$, $C_0=\frac{1}{4}zN(\frac{1}{2}\phi_{AA}+\frac{1}{2}\phi_{BB}+\phi_{AB})$, $N$ is the total number of sites, $z$ is the coordination number. Omitting the constant and external-field terms by assumption $\phi_{AA}=\phi_{BB}$ we leave the only term:
\begin{equation}
\mathcal{H}=-J\sum_{\langle i,j\rangle}S_i S_j. \label{eq:ising}
\end{equation}
If $J$ is positive, similar atoms tend to cluster together just as parallel spins are favoured in a ferromagnet. If $J$ is negative, there is a tendency to form unlike pairs, as in an antiferromagnet. If $J$ is zero, the system will be perfectly disordered. In the high temperature regime the Ising model, despite its simplicity, allows both qualitative and quantitative description of various orderings and phase transitions in binary systems~\cite{Binder1981,Arzhnikov1996,Muller2005}.

Most commonly SRO is described by the Warren--Cowley (WC) parameter~\cite{Cowley1950,Warren1969} which is the pair-correlation function:
\begin{equation}
 \alpha = 1 - \frac{P_{AB}}{x}, \label{eq:cowley}
\end{equation}
where $P_{AB}$ is the probability of a $B$ atom being found at a site nearest to the $A$ atom, $x$ is the concentration of $A$ atoms. The sign of $\alpha$ coincides with the sign of $J$, $\alpha=0$ corresponds to perfect disorder.

\subsection{\label{sub:MC}Monte Carlo simulation}
Similarity with the Ising model suggests that SRO of a finite system can be effectively studied by Monte Carlo (MC) simulation~\cite{Scholten1985}. There are the following differences from the common Metropolis--Hastings algorithm for the magnetic Ising model~\cite{Newman1999}: the alloy concentration (net magnetization in terms of magnetism) is set at the system initialization and remains fixed during simulation; not a spin-flip on site, but a change from $AB$ to $BA$ bond is made at every MC step.

We consider the system of $N=L\times L$ sites with $L=10$ and periodic boundary conditions. We start with a random distribution of atoms (spins) with the required concentration. The concentrations considered are in the range from 0.05 to 0.95 with a step of 0.05. Each trial to change the neighboring atoms $A$ and $B$ (spins $\uparrow$ and $\downarrow$) is accepted according to the Metropolis prescription (Kawasaki algorithm~\cite{Kawasaki1966}). To reach the equilibrium state we take $10^3$ MC steps for all the concentrations studied, where one MC step consists of $N$ single-site updates. Then, we take $10^{7}$ MC steps to collect the data set $\mathcal{D}$ of $10^5$ samples over which averaging will occur. The large size of the data set allows us to avoid overfitting problem and using regularization in the further training of the neural network~\cite{Torlai2016}.

Holding the concentration can be interpreted as an external field and implies the absence of phase transition in the ferromagnetic (FM) case: there is long-range order (LRO) for any value of $J$ on the square lattice. At the same time, at $J>J_c$ the system undergoes phase separation~\cite{Baxter1982}. The case of antiferromagnetic (AFM) interaction is much more complex: the model undergoes a phase transition from the ordered AFM to disordered paramagnetic (PM) state even in the presence of an external field~\cite{Binder1980}. There is no exact analytical results for the critical values and numerical simulations are also problematic~\cite{Binder1980,Lourenco2016}.

We restrict ourselves by the consideration of FM $J=+0.2$ and AFM $J=-0.2$ interactions (it is assumed that the temperature $T=1$). The magnitude of $J$ is chosen to stay in the single-phase region and to ensure that in the case of AFM interaction the system is in PM state far from phase transition for any alloy concentration. So we avoid problems with phase separation and AFM LRO, and at the same time can study two essentially different cases with the standard Metropolis MC algorithm on a relatively small lattice. Omitting the possibility of phase separation does not significantly limit the applicability of our approach, because many real alloys with short-range order have broad single-phase concentration ranges (see e.~g.~Refs.~\cite{Arzhnikov2001,Arzhnikov1996}), and the basic concept of the novel high-entropy alloys requires a single-phase state~\cite{Kozak2015,Steurer2020}. The case of phase separation needs a more detailed consideration (as it was done for non-ergodic systems in Ref.~\cite{Westerhout2020}.)

\begin{figure}[h]
\includegraphics[width=0.5\textwidth]{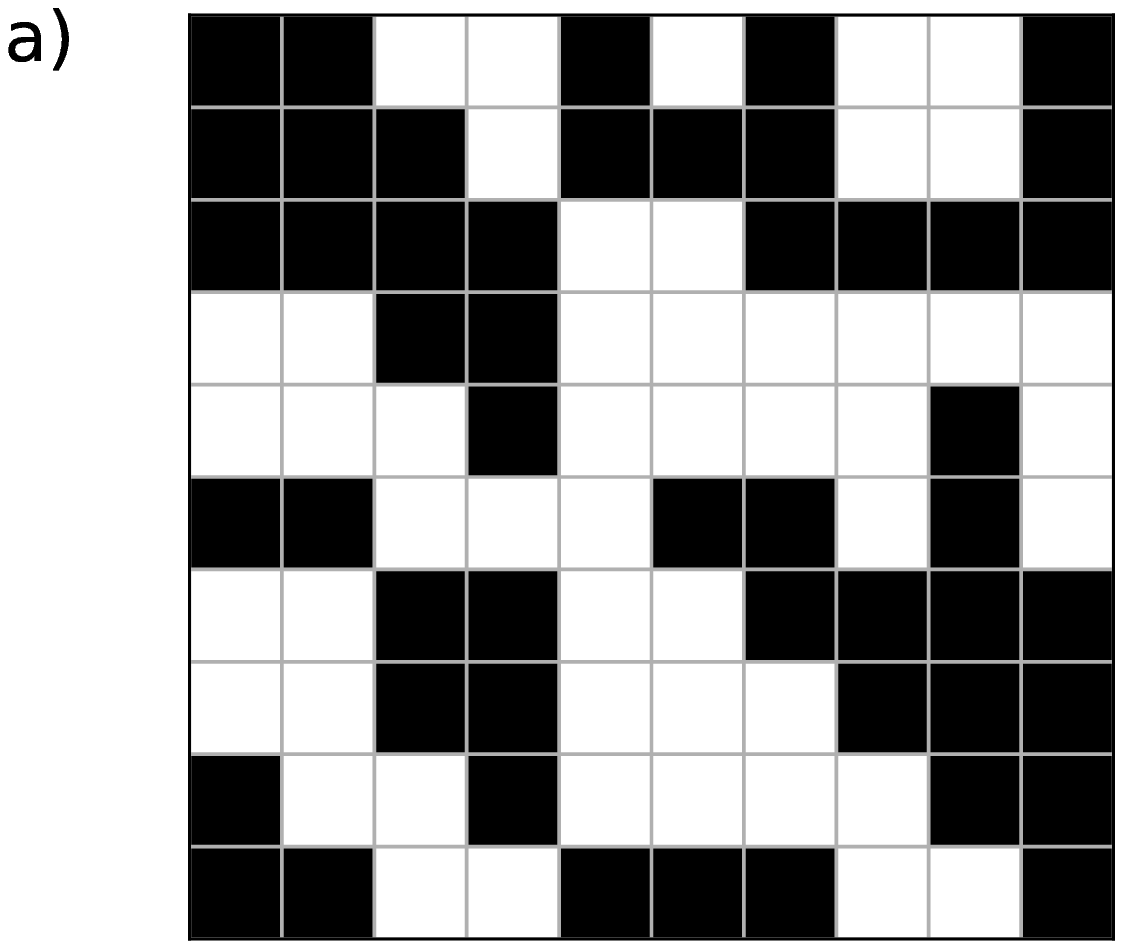} 
\includegraphics[width=0.5\textwidth]{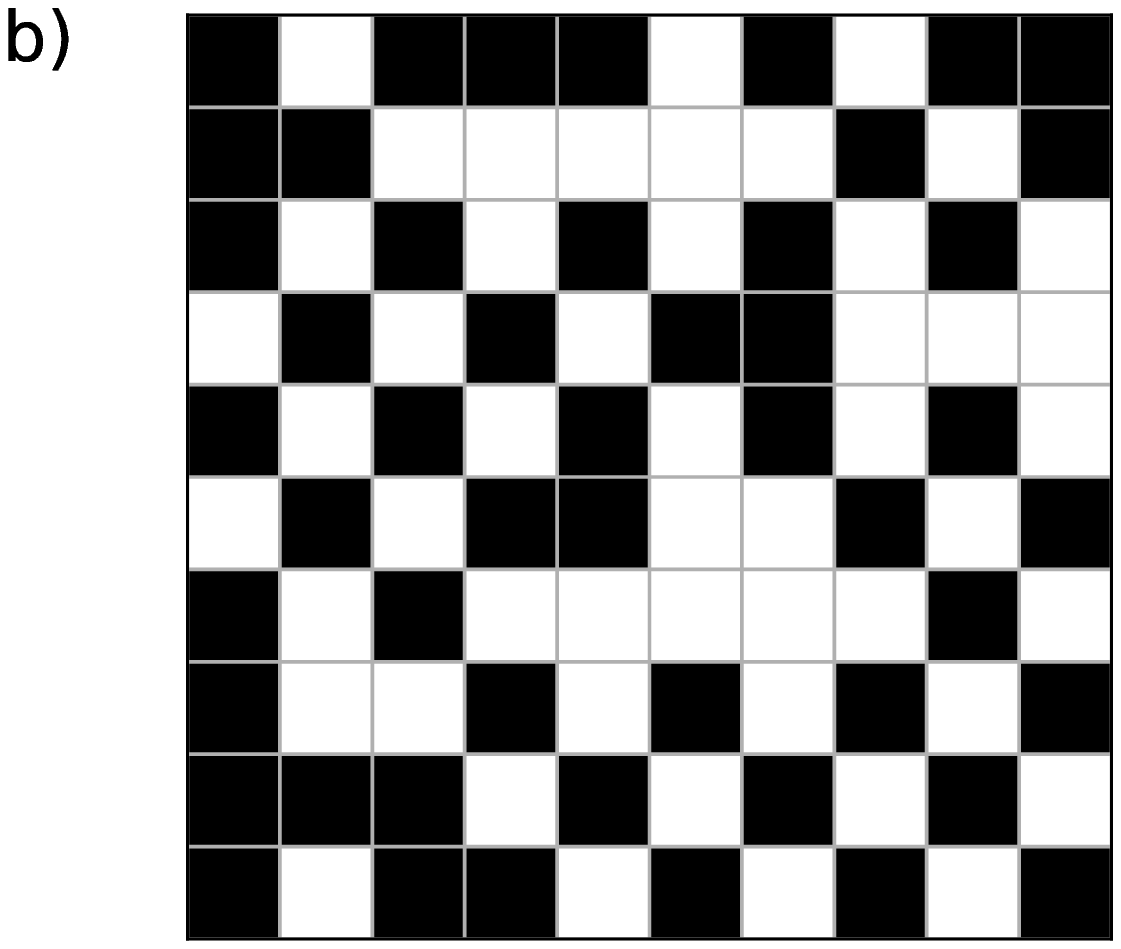}
\caption{\label{fig:samples}Examples of atoms distribution obtained by MC simulations for $J=+0.2$ (a) and $J=-0.2$ (b).}
\end{figure}

For both the FM and AFM cases we run MC simulations and collect data sets $\mathcal{D}$, over which averaging is performed. Fig.~\ref{fig:samples} shows some random samples from data sets $\mathcal{D}$ for $x=0.5$ with $J=+0.2$ and $J=-0.2$. It can be clearly seen how the atoms of the same type tend to cluster together for positive $J$ and how the atoms of different types tend to occupy neighboring sites for negative $J$. The calculated average WC parameters $\langle \alpha \rangle$ reach their extreme values at $x=0.5$ and  tend to zero at boundaries because SRO is meaningless when all of the atoms are equal. 
 
\subsection{\label{sub:RBM}Restricted Boltzmann machine}
\begin{figure}[h]
\includegraphics[width=0.6\textwidth]{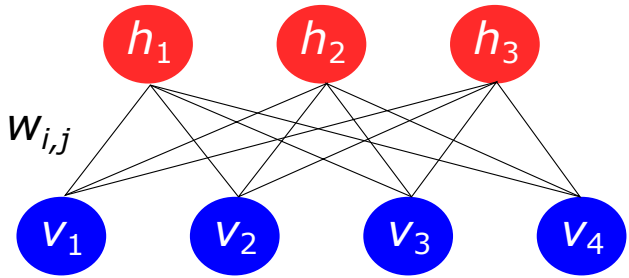}
\caption{\label{fig:RBM}Restricted Boltzmann machine architecture}
\end{figure}

The full Boltzmann machine, a stochastic energy-based neural network, consists of two types of neurons (nodes): visible and hidden~\cite{Ackley1985}. Each node is binary and connected to other nodes with some weight. If the connections between neurons of the same type are forbidden, the model is called the restricted Boltzmann machine~\cite{Smolensky1986}. The architecture of RBM is depicted in Fig.~\ref{fig:RBM}: all neurons are arranged in visible ($i=1\dots n$) and hidden ($j=1\dots m$) layers interconnected with weights $w_{i,j}$. Together with biases $a_i$ and $b_j$, the weights constitute the RBM parameter set denoted by $\theta$. These parameters determine the {\it energy} of a given configuration $(v,h)$:
\begin{equation}\label{eq:RBM_energy}
E(v,h)=-\sum_i a_i v_i-\sum_j b_j h_j - \sum_{i,j} v_i w_{i,j} h_j.
\end{equation}
The probability distribution over hidden and visible vectors is defined by Boltzmann distribution:
\begin{equation}\label{eq:prob}
P(v,h)=\frac{1}{Z}e^{-E(v,h)},
\end{equation}
where $Z$ has the meaning of the partition function.

RBMs gained wide popularity in the 2000s, because some efficient training algorithms were developed by Hinton~\cite{Hinton2002}.
Once RBM is trained, which means that the parameter set $\theta$ is found, a new data set $\mathcal{S}$ can be generated by the block Gibbs sampling procedure. Successful training provides similarity between the real physical probability distribution and the probability distribution of the generated data set. The method of RBM training is described in detail in Ref.~\cite{Hinton2012}. The main idea is to use the so-called contrastive divergence approximation with $k$ steps (CD-$k$) to perform stochastic gradient descent procedure for the negative log-likelihood function. Contrastive divergence procedure becomes possible due to bipartite structure of RBM, which enables the block Gibbs sampling. The hidden layer can be completely obtained from the known visible layer, and vice versa, by the conditional probabilities:
\begin{eqnarray}
p(h_j=1|v)=\sigma\left(b_j+\sum_i w_{i,j} v_i \right), \label{eq:ph}\\
p(v_i=1|h)=\sigma\left(a_i+\sum_j w_{i,j} h_j \right). \label{eq:pv}
\end{eqnarray} 
To generate a new data set by trained RBM one can initialize visible layer by random values $v_0$ and run Gibbs sampling procedure i.~e. to compute hidden layer $h_0$, then new visible layer $v_1$, then new hidden layer $h_1$ and so on. After some equilibration period (like in usual MC method) the generated visible units can be collected to the data set $\mathcal{S}$, which is expected to obey the same probability distribution as the data set $\mathcal{D}$. Averaging over $\mathcal{S}$ allows to calculate the observable physical values like mean energy or heat capacity.

\section{\label{sec:results}Results}

In order to produce the training data set $\mathcal{D}$, we use the standard Markov chain MC technique described in Sec.~\ref{sub:MC}. The importance sampling yields $10^5$ independent spin configurations for $J=+0.2$ and $J=-0.2$ for each concentration $x \in [0.05;0.95]$ in steps of $\Delta x = 0.05$.

\begin{figure}[h]
\includegraphics[width=0.8\textwidth]{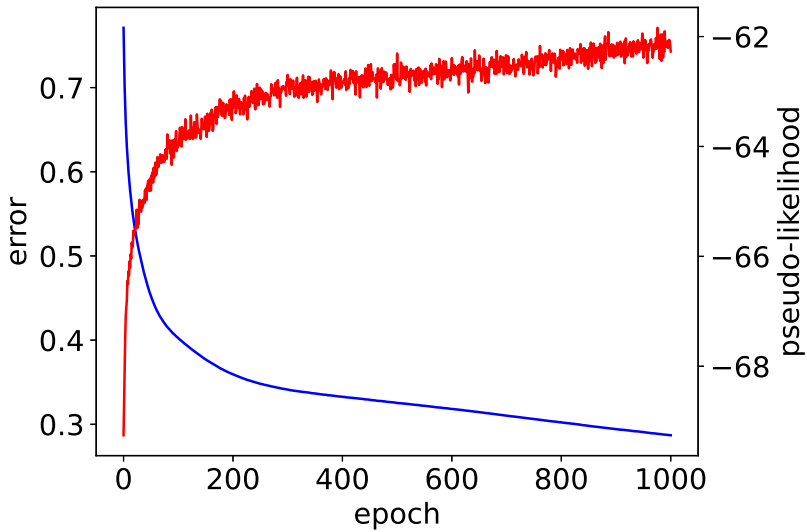}
\caption{\label{fig:training}Monitoring the RBM training process. Reconstruction error (blue) and pseudo-likelihood (red).}
\end{figure}

Preliminary examinations have given us the following optimal RBM parameters, which were used further for training: the number of hidden units is equal to the number of visible units $n_h=n_v=100$, contrastive divergence with the only one step (CD-$1$), learning rate $\eta=0.01$, minibatch size 100, the number of learning epochs 1000 (definitions of training parameters can be found in Appendix). RBMs with these parameters were trained on data set $\mathcal{D}$ at each concentration of the studied range. In order to illustrate the learning process, we present typical graph of decreasing reconstruction error (mean square deviation between original and reconstructed data~\cite{Hinton2012}) and increasing pseudo-likelihood (approximated value of the likelihood function~\cite{Hyvarinen2006}) during training~(Fig.~\ref{fig:training}).

\begin{figure}[h]
\includegraphics[width=0.8\textwidth]{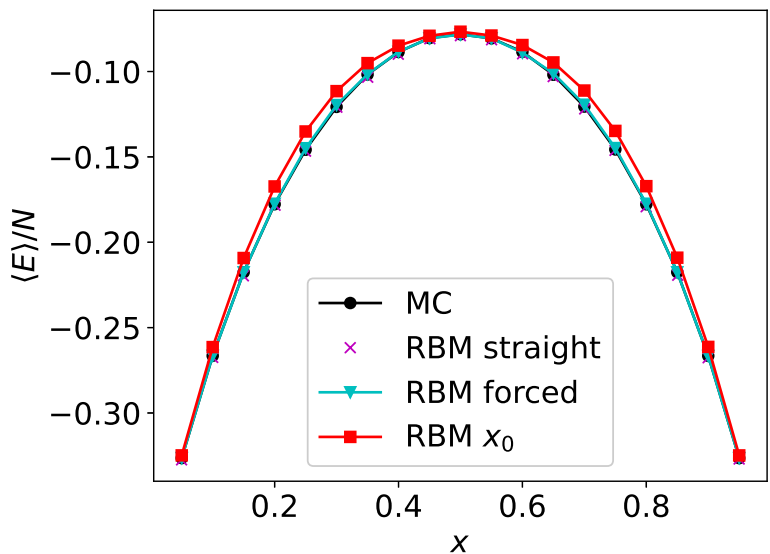}
\caption{\label{fig:energyFM_full}Energy per site for FM interaction. Black solid line denotes MC simulation, cross markers are straight RBM, cyan triangles are forced RBM, red squares are RBM trained on $x_0=0.5$ data set.}
\end{figure}

\begin{figure}[h]
\includegraphics[width=0.8\textwidth]{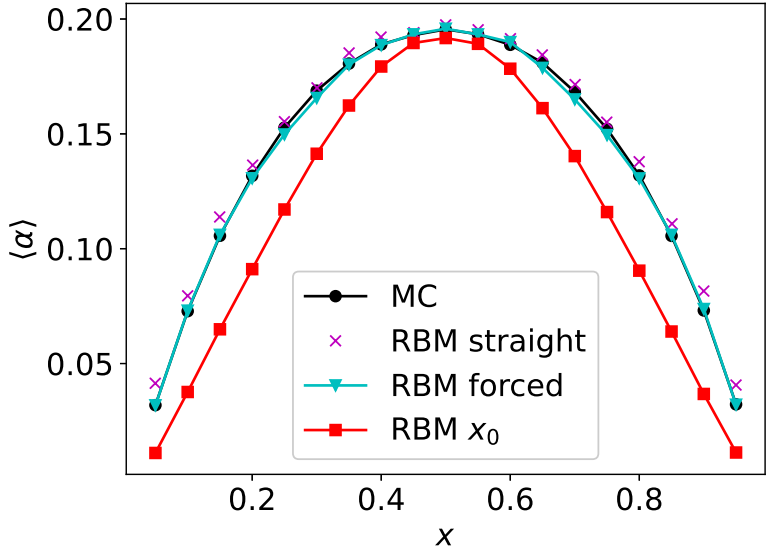}
\caption{\label{fig:wcFM_full}Warren--Cowley parameter for FM interaction. Black solid line denotes MC simulation, cross markers are straight RBM, cyan triangles are forced RBM, red squares are RBM trained on $x_0=0.5$ data set.}
\end{figure}

\begin{figure}[h]
\includegraphics[width=0.8\textwidth]{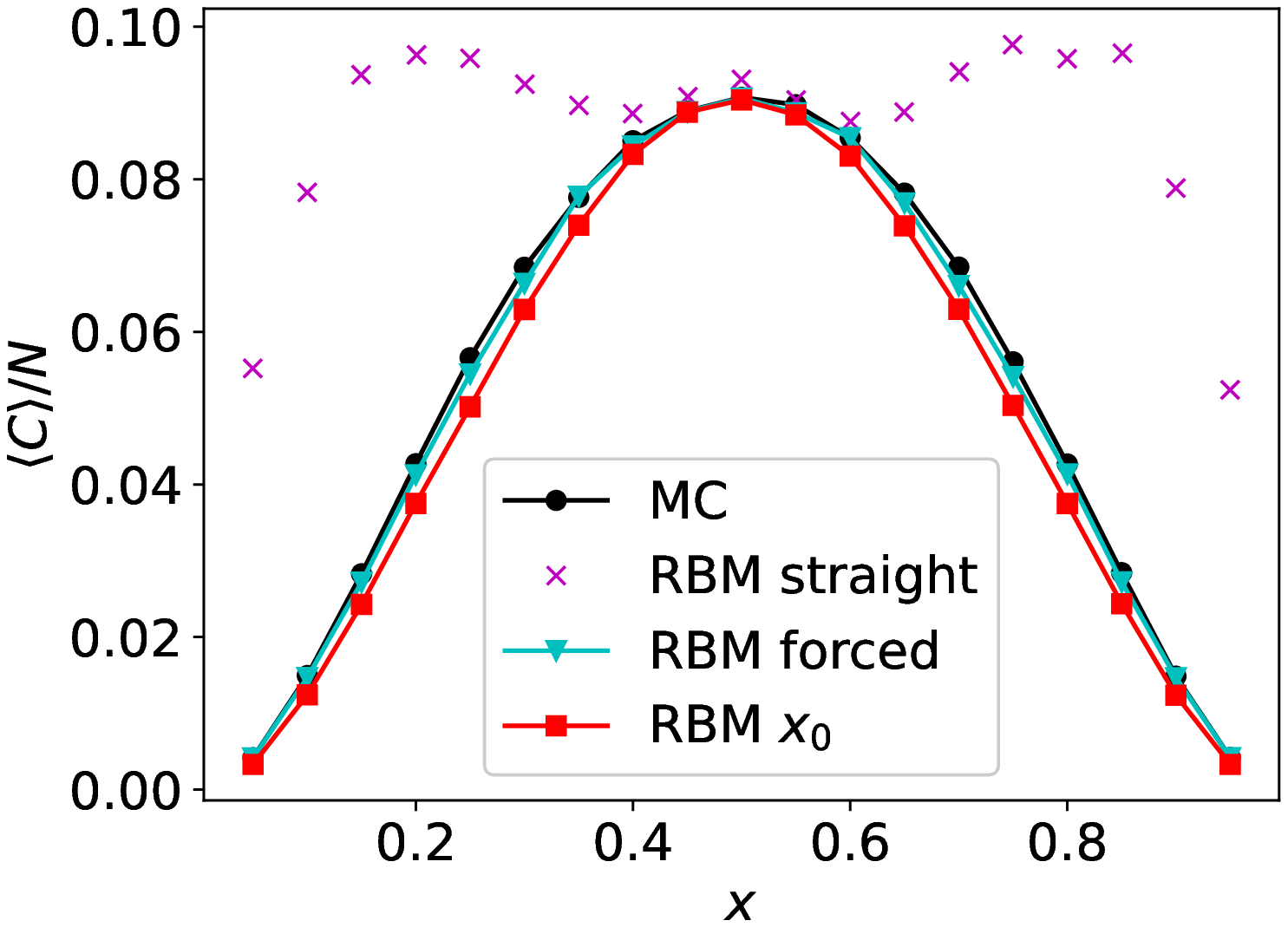}
\caption{\label{fig:capacityFM_full}Heat capacity per site for FM interaction. Black solid line denotes MC simulation, cross markers are straight RBM, cyan triangles are forced RBM, red squares are RBM trained on $x_0=0.5$ data set.}
\end{figure}

At the first stage we address the question how well can RBM reproduce the observable physical quantities known from MC simulations. In order to answer it, we generate a new data set $\mathcal{S}$ of $10^5$ samples at each concentration by the corresponding RBMs and calculate the average observables. 100 Gibbs steps were used to reach the equilibrium state and then the samples were recorded in $\mathcal{S}$ every 50 steps.  In Figs.~\ref{fig:energyFM_full}-\ref{fig:capacityFM_full} computed by such a way energy $\langle E \rangle$, WC parameter $\langle \alpha \rangle$ and heat capacity $\langle C \rangle = \left( \langle E^2 \rangle - \langle E \rangle ^2\right)/T^2$ are depicted as 'straight RBM' and compared with the 'exact' values obtained by MC simulations for FM interaction. While the WC parameter and the energy show good agreement, the heat capacity is not satisfactory at all. This is explained by strong sensitivity of long-range correlations to the alloy concentration which is actually is variable in each data set. The reason is that the Gibbs sampling scheme being fundamentally stochastic procedure gives the concentration $x_0$ (at which RBM was trained) only on average, but the concentrations of each individual sample of $\mathcal{S}$ is fluctuating around $x_0$. These fluctuations result in fluctuations of the heat capacity. 

\begin{figure}[h]
\includegraphics[width=0.8\textwidth]{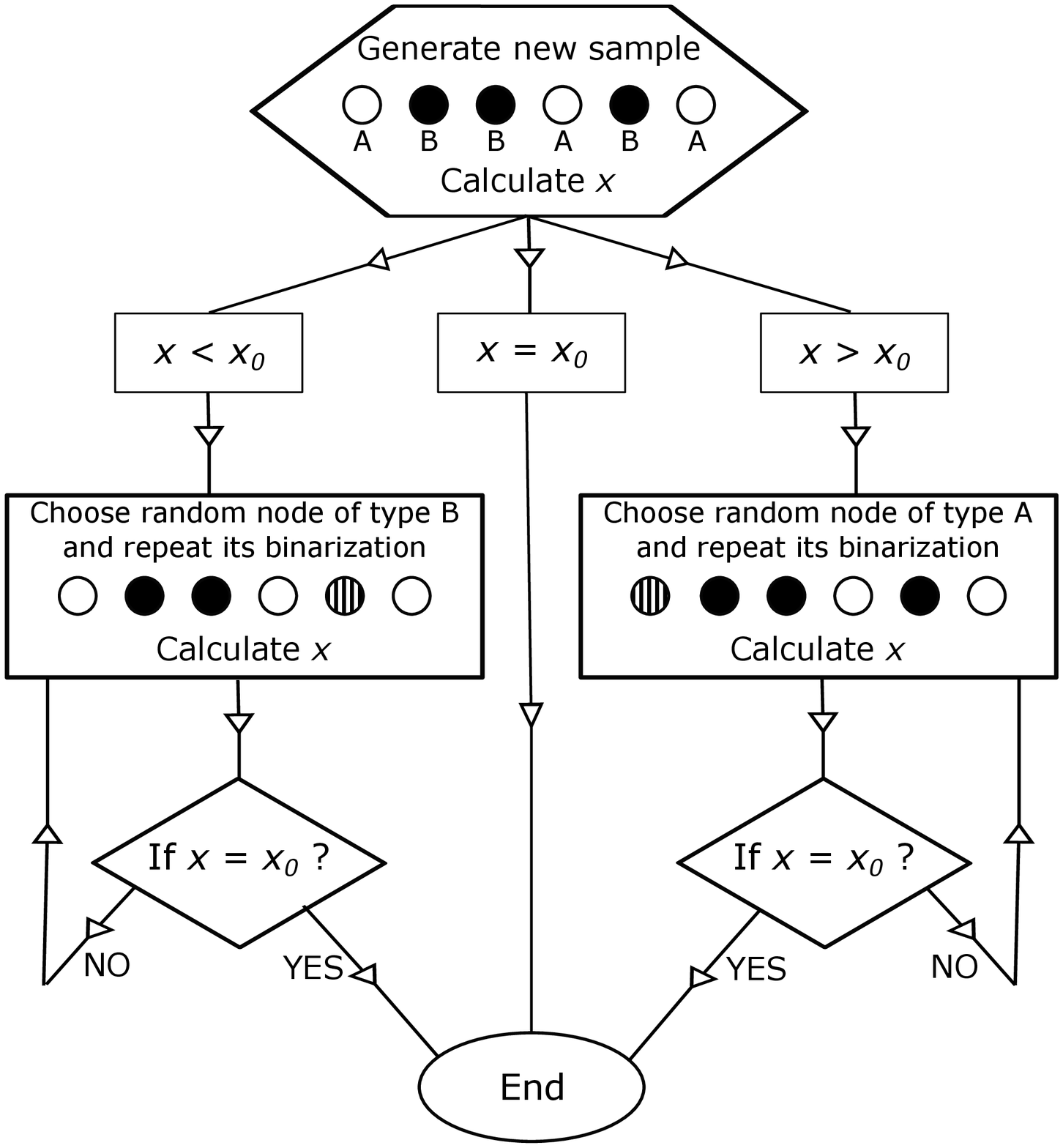}
\caption{\label{fig:block}Flowchart of the sample generation with the forced concentration $x_0$.}
\end{figure}

\begin{figure}[h]
\includegraphics[width=0.8\textwidth]{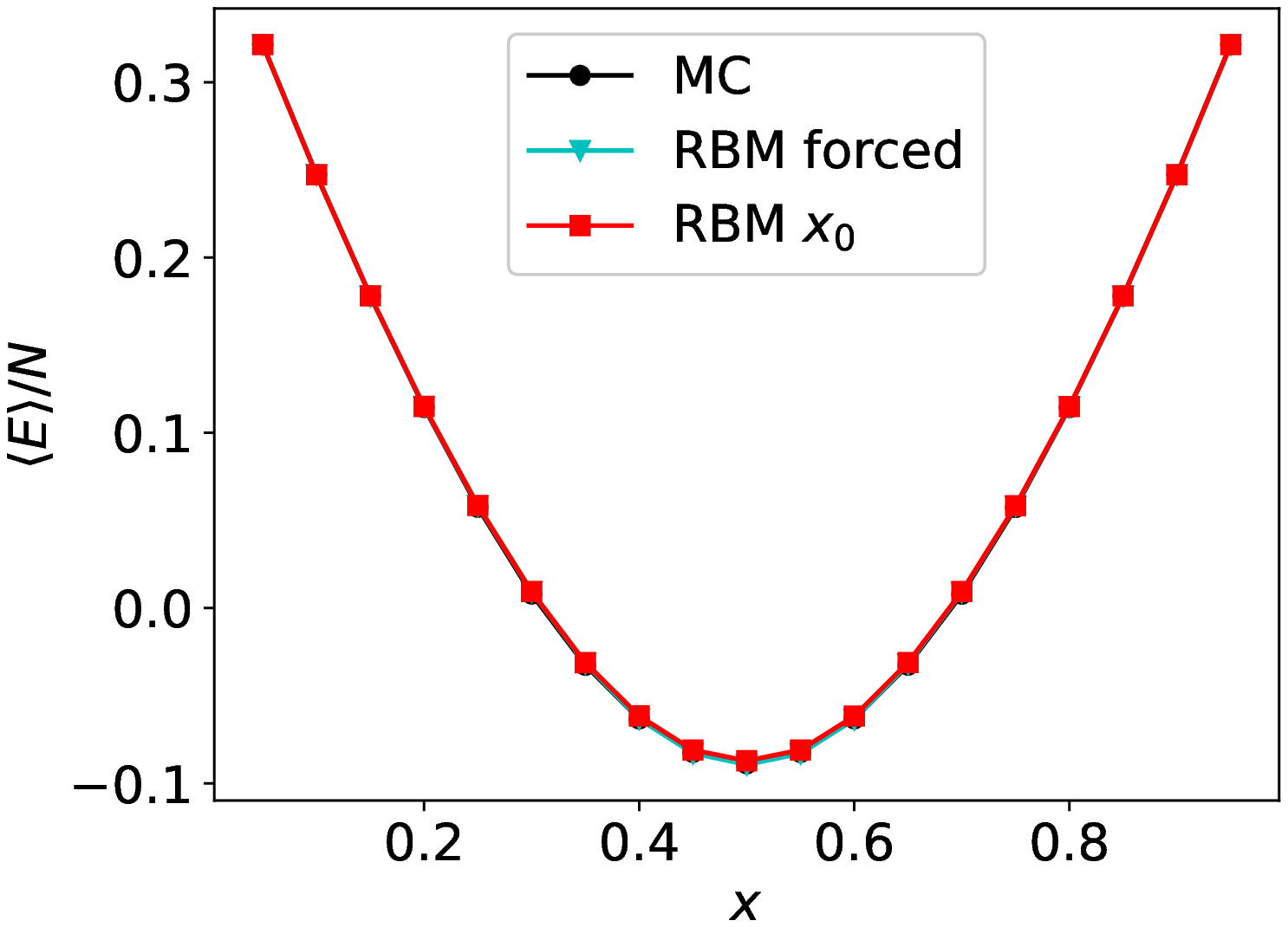}
\caption{\label{fig:energyAFM02}Energy per site for AFM interaction. Black solid line denotes MC simulation, cyan triangles are forced RBM, red squares are RBM trained on $x_0=0.5$ data set.}
\end{figure}

\begin{figure}[h]
\includegraphics[width=0.8\textwidth]{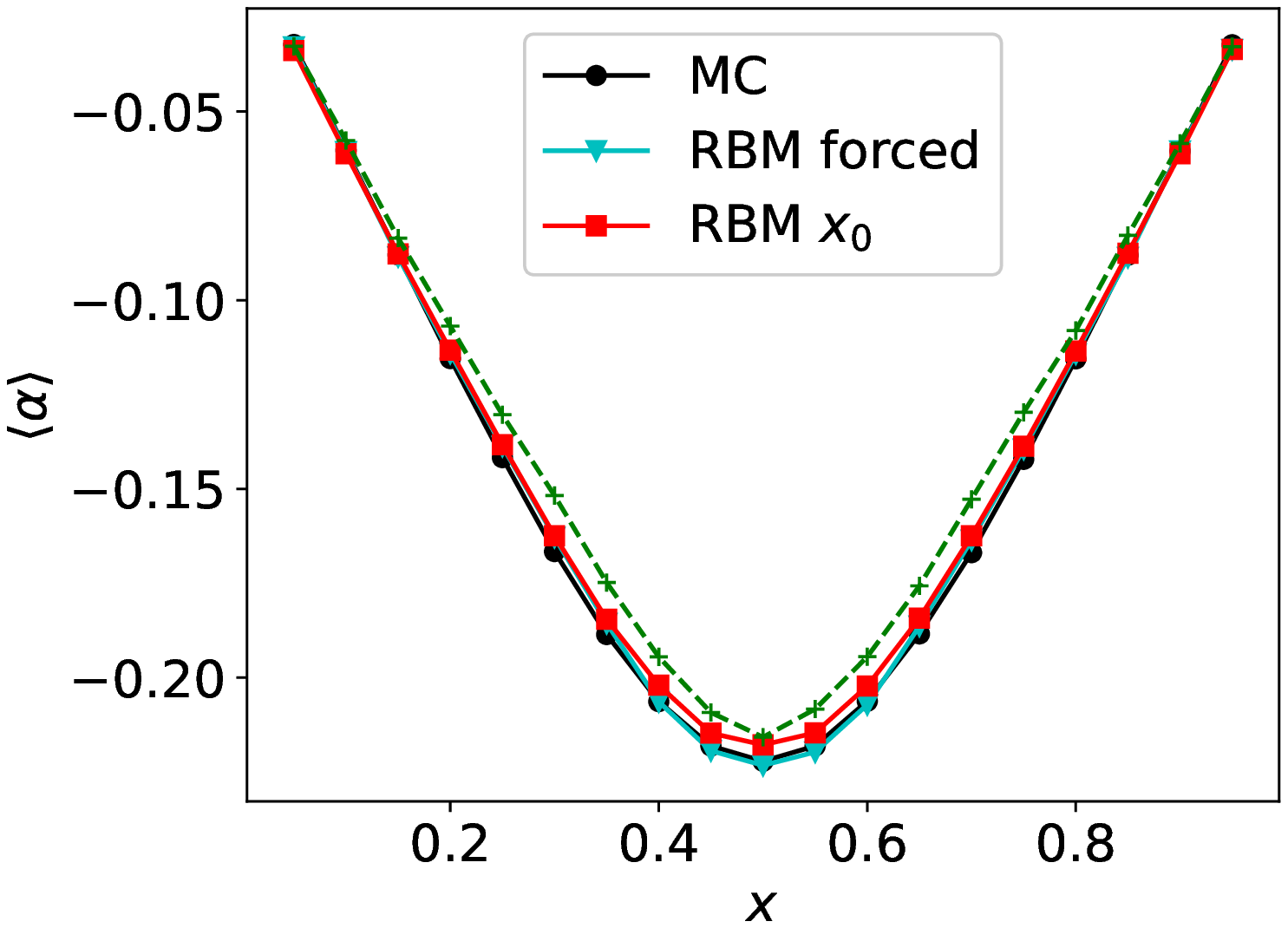}
\caption{\label{fig:wcAFM02}Warren--Cowley parameter for AFM interaction. Black solid line denotes MC simulation, cyan triangles are forced RBM, red squares are RBM trained on $x_0=0.5$ data set. Green dashed line is calculated by RBM trained on reduced data set.}
\end{figure}

\begin{figure}[h]
\includegraphics[width=0.8\textwidth]{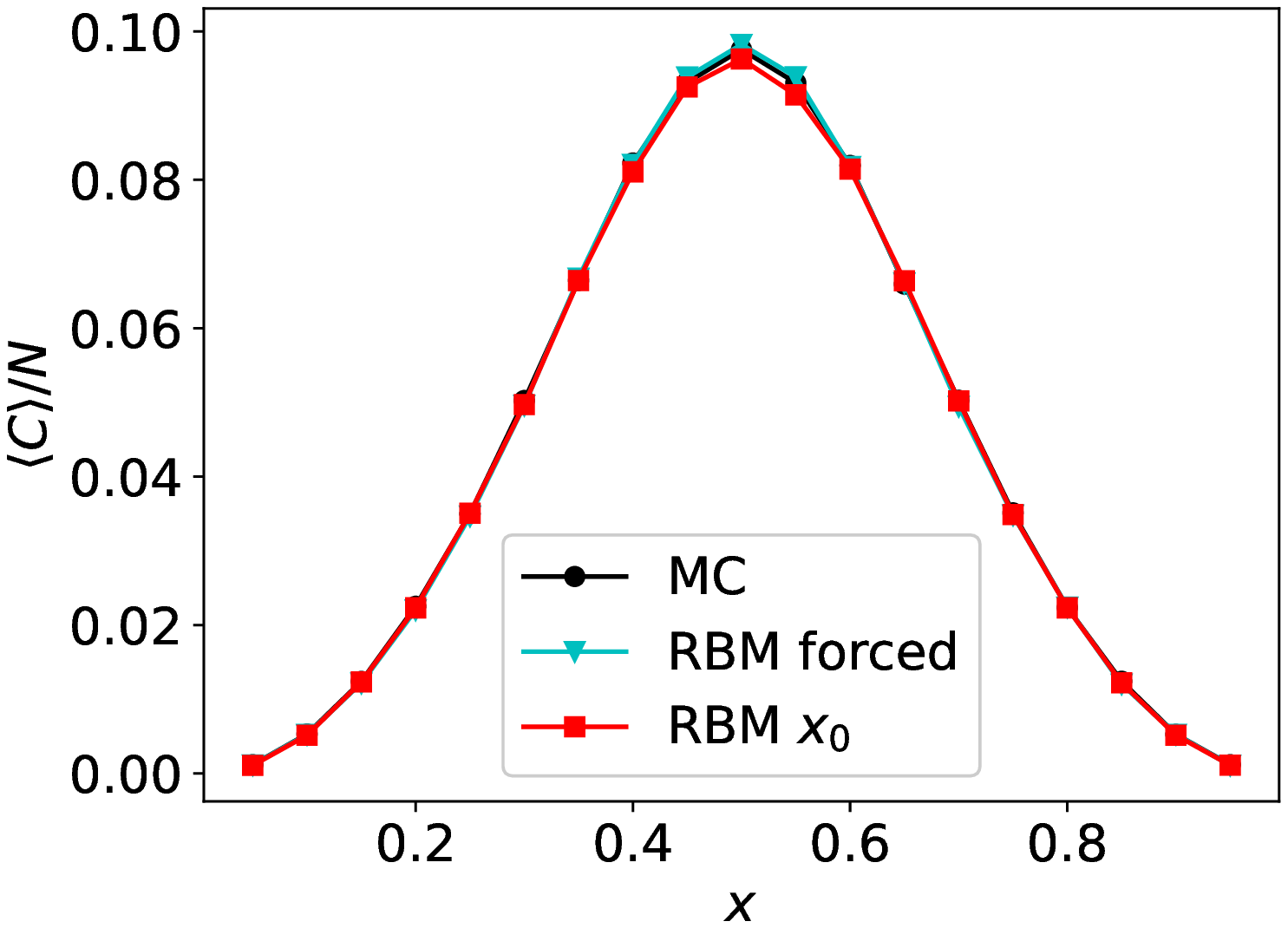}
\caption{\label{fig:capacityAFM02}Heat capacity per site for AFM interaction. Black solid line denotes MC simulation, cyan triangles are forced RBM, red squares are RBM trained on $x_0=0.5$ data set.}
\end{figure}

To overcome this difficulty, we propose an  algorithm for forcing the concentration to have the required value $x_0$. A flowchart of the algorithm is depicted in Fig.~\ref{fig:block}. Every time a new sample (visible layer) has been generated, its concentration is calculated immediately. If the concentration is equal to the required $x_0$, we accept this sample. If the concentration is lower than $x_0$, we need to reduce the number of excess $B$ atoms. To do this, we choose a random node $k$ of type $B$ and re-make its binarization by comparing $p(v_k=1|h)$ (\ref{eq:pv}) with a random number from 0 to 1. If the atom type changes but the concentration is still not equal to $x_0$ we repeat the procedure of random atom selection and re-binarization until the required concentration $x_0$ is reached. Similar steps are taken in the case of $x>x_0$. Finally, we record the layer with the concentration $x_0$ in the dataset. So we achieve that all the samples have the same concentration $x_0$. Averaging over the data sampled by RBM with the described forcing procedure gives the values presented as 'RBM forced' in Figs.~\ref{fig:energyFM_full}-\ref{fig:capacityFM_full} for FM interaction and Figs.~\ref{fig:energyAFM02}-\ref{fig:capacityAFM02} for AFM interaction. As can be seen all the observables now agree much better with MC data.

The goal of the second stage of our research is to check whether it is possible to access the information about SRO at some arbitrary concentration by RBM trained at completely different concentration. In order to do it one should have a possibility to generate samples with arbitrary concentration. We assumed that the algorithm described by flowchart (Fig.~\ref{fig:block}) can be applied to solve this problem. Indeed, this procedure does not contain any restrictions for the concentration $x_0$: stochastic nature of RBM allows for forcing any concentration $x'$ in sample with a sufficient number of repetitions of the flowchart steps. The only limitation is that in the case of an excessively large difference between the desirable concentration $x'$ and the concentration of the training data set $x_0$ ($\Delta x=x'-x_0$) the duration of such an algorithm may turn out to be too long. To make it easier to achieve the required concentration, we modify the probability definition (\ref{eq:pv}) by adding to it the difference $\Delta x$. Making this we shift the average concentration in a sample generated by RBM to the side of $x'$. As a result we have an efficient and fast method to force the alloy concentration in the generated data set $\mathcal{S}$ to have an arbitrary value. 

With the help of the described algorithm we use RBM trained at $x_0=0.5$ to generate data for all the concentrations from 0 to 1 with the step $0.05$. $10^5$ samples are prepared for each concentration and the average observables are computed. The results are presented as 'RBM $x_0$'  in Figs.~\ref{fig:energyFM_full}-\ref{fig:capacityFM_full} for FM interaction and in Figs.~\ref{fig:energyAFM02}-\ref{fig:capacityAFM02} for AFM interaction. A comparison with the 'exact' MC data leads to an impressive conclusion: RBM trained at one concentration satisfactorily reproduces both qualitative and quantitative character of SRO and other observables at any concentration, including those far from original $x_0$.

It is important to note that the conclusions obtained are stable in relation to changes in the training parameters. In particular, reducing the training data set ($\mathcal{D}$ of $10^5$ configurations is needed to accurately calculate the observables, but is probably redundant for training), preserves the predictive power of RBM. The WC parameter calculated with a machine trained at a concentration of $x_0=0.5$ on a data set of $10^3$ samples for the AFM case is shown in Fig.~\ref{fig:wcAFM02} by dashed line. It can be seen that it still agrees well with the exact data.

\section{\label{sec:conclusion}Conclusions}

We demonstrate the predictive power of generative neural networks for the systems exhibiting short-range order. The algorithm is proposed that allows using a machine trained on some alloy concentration to calculate the order parameters at any other alloy concentration. Compared to the regular Metropolis--Hastings algorithm, in which MC trials for each site are performed strictly sequentially, Gibbs sampling, thanks to the RBM architecture, allows sampling all the nodes simultaneously, which is much more efficient and opens the way to parallel computations.

Applicability of our algorithm appears to be remarkably wide, it is suitable for the prediction of the SRO and thermodynamic properties in the cases of positive and negative effective interactions, the ordered and disordered phases, close and far from the original concentration. This may be explained by the inherent ability of the neural networks to recognize the patterns underlying the data, combined with an exceptional ability of RBM, as a stochastic generative network, to reproduce these patterns in the sampled data in a controlled manner. 

The proposed method is approved for the two-component system but can be naturally extended to the case of multicomponent alloys (RBM extension to non-binary units is considered in Ref.~\cite{Hinton2012}). This can be particularly useful in the context of the high-entropy alloys investigation, novel materials with exceptional mechanical properties. They are consisted of five or more elements and their characteristics are very sensitive to local chemical environment. Monte Carlo simulations for such systems (see e.~g.~\cite{Schonfeld2019}) are very expensive in terms of computational time and the search for the optimal composition by trying all possible concentrations is problematic. RBM trained on some samples taken from MC simulations would shed light on inaccessible areas of concentration, spending relatively little computing resources.

Another promising application is to find a local atomic environment in alloys based on experimental data for a single concentration. For example, if we have the probabilities of the clusters with different atomic configurations from the experiment, we can use MC approach to generate a data set where each sample is ordered accordingly. Using our algorithm, we can train RBM on this data set and predict the local atomic configurations, and hence the other properties, e.~g. hyperfine field or local magnetization~\cite{Arzhnikov1996}, for the concentrations other than the experimental one.

\section*{\label{sec:acknowledgments}Acknowledgments}
This study was supported by the financing program AAAA-A16-116021010082-8.

\section*{\label{sec:appendix}Appendix: RBM training}
Unsupervised RBM training implies learning an unknown distribution $P(v)$ based on sample data. Specifically for RBM it means adjusting the parameters $\theta=\{w_{i,j},a_i,b_j\}$ of the probability distribution $P(v|\theta)$ that maximize the likelihood of the training data set $\mathcal{D}=\{v_1..v_n\}$. Maximizing the likelihood $\mathcal{L}(\theta|\mathcal{D})$ corresponds to minimizing the distance between the unknown distribution $P_{data}$ underlying $\mathcal{D}$ and the model distribution $P_{model}$ in terms of the Kullback-Leibler divergence (KL divergence), which is given by:
\begin{equation}\label{eq:KL}
\mathrm{KL}(P_{data}||P_{model})=\sum_i P_{data}(v_i)\ln P_{data}(v_i)-\sum_i P_{model}(v_i)\ln P_{model}(v_i).
\end{equation}
Differentiating this we can solve the optimization problem and deduce an update rules for the gradient ascent on the log-likelihood~\cite{Hinton2002}:
\begin{eqnarray}\label{eq:update}
w_{i,j}^{(t+1)}=w_{i,j}^{(t)}+\eta\left(\langle v_i h_j\rangle_{data}-\langle v_i h_j \rangle_{model}\right) \\
a_i^{(t+1)}=a_i^{(t)}+\eta\left(\langle v_i \rangle_{data}-\langle v_i \rangle_{model}\right) \\
b_j^{(t+1)}=b_j^{(t)}+\eta\left(\langle h_j\rangle_{data}-\langle h_j \rangle_{model}\right).
\end{eqnarray}
$\eta$ is the learning rate.

Average $\langle\ldots\rangle_{model}$ values can be obtained by Gibbs sampling chain:
\begin{equation}\label{eq:gibbs}
v^0\rightarrow h^0\rightarrow v^1 \rightarrow h^1\rightarrow \ldots \rightarrow v^k  \rightarrow h^k
\end{equation}
using conditional probabilities (\ref{eq:ph})-(\ref{eq:pv}) until the stationary distribution is reached. The idea of RBM training is to limit the chain (\ref{eq:gibbs}) to a few $k$-steps, which turned out to be sufficient for effective learning. This is so called contrastive divergence (CD-$k$) method, and in practice $k=1$ is usually used~\cite{Hinton2002}. Then the update rules look like this:
\begin{equation}\label{eq:CD}
w_{i,j}^{(t+1)}=w_{i,j}^{(t)}+\eta\left(v_i^0 h_j^0-v_i^1 h_j^1\right).
\end{equation}
To speed up the training procedure the data set $\mathcal{D}$ is split into minibatches of small size and the update of weights and biases is done after each minibatch and not after the entire set.

Starting from zero biases $a_i$, $b_j$ and random weights $w_{i,j}$ normally distributed around zero we run an iterative procedure, each step of which corresponds to one learning epoch.
The above algorithm is the standard way for RBM training and has proven to be effective in numerous applications.

\bibliographystyle{elsarticle-num}
\bibliography{RBM}

\end{document}